# The Impossibility of the Almost Pareto Principles

First version: 1st October 2024,

This version: 22th November 2024.

Norihito Sakamoto[*]

## Abstract

This study proposes a new efficiency requirement, a *minimal almost weak Pareto* principle, which says that $x$ is socially better than $y$ whenever the only one individual never prefers $y$ to $x$, and all the others prefers $x$ to $y$. Then, I show that even if the Pareto principle is modified into this harmless form, that seems sufficiently acceptable in the setting of social choice with variable population sizes or incomplete preferences, it violates acyclicity. Furthermore, it is shown that under this framework, a modified Pareto indifference and usual weak Pareto are inconsistent. These results are serious because they have a wide range of applications, not only to population economics and intergenerational equity analysis, but also to welfare evaluations of incomplete preferences and multi-dimensional well-being. In order to solve these problems, it is necessary to impose very strong assumptions on various contexts of social choice problems.

[*] Tokyo university of science, n-sakamoto@rs.tus.ac.jp
I thank Gustaf Arrhenius, Geir B. Asheim, Timothy Campbell, Marc Fleurbaey, Karim Jebari, Eric Mohlin, Jake Nebel, Daniel Ramöler, and Stéphane Zuber for their helpful comments and suggestions which greatly contribute to revising the paper. Also, I appreciate financial support from JSPS KAKENHI (Grant number: 20KK0036).

# 1. Introduction

The Pareto principle has been apparently a dominant and incontrovertible axiom as a concept of efficiency in economics for about 100 years. However, since the 1960s, it has been pointed out that this sacred dogma contradicts competing principles that seem as a basis of a liberal and democratic society. For example, Sen (1970) showed that a combination of minimal liberalism and weak Pareto violates acyclicity. In the context of evaluating infinite utility streams, Diamond (1965) pointed out the incompatibility among very standard axioms: anonymity, strong Pareto, and indifference transitivity. Kaplow and Shavell (2001) proved that any non-welfarist social welfare function cannot satisfy both weak Pareto and continuity.[1] These impossibility results show fundamental conflicts between efficiency requirements and other important values in our society.

While the Pareto principle is usually a normative concept of efficiency in a fixed population model, in order to compare well-being profiles with different population sizes and generations, a social planner needs to appropriately extend this efficiency concept for problems of population ethics and population economics.[2] However, the folk knowledge suggests that a natural extension of the Pareto principle may yield a cycle without imposing any other requirements (Temkin 1987; Golosov et al. 2007). In fact, Temkin (1987) showed that an extension of the weak Pareto principle violates acyclicity in the following simple example.

***Example*** (Temkin 1987). Consider three potential individuals 1, 2, and 3, and three social states *u*, *v*,

---

[1] Kaplow and Shavell (2001) assume that non-welfarist assessment is equivalent to a violation of Pareto indifference. Generally, under mild assumptions including continuity, all the Pareto principles are equivalent (Suzumura 2001). Hence, note that a part of results in Weymark (1993) and Suzumura (2001) are formally equivalent to Kaplow and Shavell's impossibility theorem. See Fleurbaey et al. (2003) for a critical review of Kaplow and Shavell's interpretation of non-welfarism and welfarism.

[2] Needless to say, the origins of economic analysis of population issues can be traced back to the works of Malthus and J. S. Mill. In response to growing concerns about population explosion and environmental issues from the 1960s, population ethics and population economics has rapidly grown since the 1980s.

and *w*. Each individual's utility is represented by a real value, with larger values being more desirable. Suppose utility levels of each individual in *u*, *v*, and *w* are given in the following figure:

|   | 1 | 2 | 3 |
|---|---|---|---|
| *u* | (2, | 1, | ∅), |
| *v* | (1, | ∅, | 2), |
| *w* | (∅, | 2, | 1). |

In the figure, ∅ means that the individual does not exist in the social state. For example, in the state *u*, individual 3 does not exist. As you can see in *u* and *v*, individual 1 prefers *u* to *v* because she receives strictly higher utility in *u*. Obviously, individual 2 does not exist in *v*, and individual 3 does not exist in *u*, so they cannot compare *u* and *v*. Since no individual prefers *v* to *u* and there is one individual who prefers *u* to *v*, the Pareto principle simply seems to imply that *u* is socially better than *v*. By a similar argument, *v* is socially better than *w*, and *w* is socially better than *u*. Thus, a cycle occurs.

Temkin's example may suggest one of unsolved puzzles in population ethics.[3] However, his example simply involves welfare comparisons in a world with three potential individuals, and this extended Pareto principle only needs a support of a just single person as a basis for Pareto comparisons (i.e. it is not based on any unanimous judgments). In this sense, it seems to lack a substantial basis for Pareto comparisons in a world with three potential individuals (there are no supporters of majority for any comparisons of states). Therefore, this paper requires that, for a fixed number of potential *n* individuals, the Pareto principle should be applied if there is the almost unanimous support of (*n*-1) individuals and the remaining one does not oppose. However, I show that even this very limited extension of the Pareto principle cannot satisfy acyclicity. Furthermore, I show that even the almost Pareto condition limited to indifferent relations is inconsistent with weak Pareto, when combined with indifference transitivity of social ranking. These results suggest that there are major problems with applying the almost unanimous Pareto principles in various contexts of social choice. Thus, the requirements of the Pareto principle, which many economists take for granted, turn out to be unacceptable unless they apply it only to the same population size or impose highly arbitrary assumptions about well-being of individual existence (a well-being level of a life worth living).

---

[3] This problem has also been known in the context of welfare evaluation in population economics. For example, Golosov et al. (2007) propose the concept of *A-Pareto* (introduced later in section 2) and point out that the A-Pareto principle violates acyclicity. Temkin himself argues that transitivity should be abandoned, and a society should allow cycles to occur in moral judgements. However, Temkin's intransitive view would be highly unacceptable in economics. Even in the field of analytical philosophy, it does not seem to be common to strongly advocate the abandonment of transitivity.

The contributions of this paper are as follows. First, the paper introduces new Pareto axioms that are based on almost unanimous judgments, which are natural extensions of the usual Pareto principle in applying it to various contexts. Second, I show that even minimal forms of the extended Pareto principle contradict the consistency condition by itself. Third, this study illustrates that the similar contradiction between the Pareto principle and acyclicity is common in various contexts (variable population, intergenerational or intertemporal aggregation problems, incomplete preferences, risky and uncertain situations, and multidimensional well-being), and argue that a fairly arbitrary assumption is required to resolve this problem. Because the Pareto principle is the most uncontroversial dogma in economics, the impossibility results of this paper seem simple but very serious.

This paper is structured as follows. Section 2 explains the basic notation and definitions. Section 3 presents two basic impossibility results and gives examples of how broad a range of contexts the impossibility results hold. Section 4 discusses strategies to avoid the impossibility results and explains that each of these requires very strong limitations. Section 5 concludes the results.

## 2. Notation and Definitions

This section explains the notation and definitions used in this paper. Let $\mathcal{N}$ be a family of possible finite sets of individuals. Although all finite population sizes are included, the actual analysis in this paper is limited to comparing societies with population sizes of *n* and (*n*-1) people with the same identity (i.e., each person has a fixed preference).[4] Each person has each preference, which can be represented by a real value for the sake of simplicity.[5] A vector of utility levels of individuals is called a utility profile, and its universal set is denoted as $U = \bigcup_{n \in \mathbb{N}} (\mathbb{R} \cup \emptyset)^n$ where the symbol $\emptyset$ represents the non-existence of individuals. A social ranking $\succcurlyeq$ is defined on the set of utility profiles, and for any $u_N, v_M \in U$, $u_N \succcurlyeq v_M$ is interpreted as $u_N$ being at least as socially good as $v_M$. The social ranking is assumed to satisfy reflexivity.[6] Also, $\succ$ and $\sim$ are the asymmetric and symmetric factors of $\succcurlyeq$, respectively.

The only requirements for social rankings[7] are the conditions of consistency and Pareto efficiency. Note that completeness and transitivity often are imposed on social rankings, but these are not required in this paper. Hence, a natural extension of Pareto is fundamentally incompatible with the consistency conditions.

---

[4] As examples of comparing well-being profiles of different population sizes with the same identities, the reader could imagine traffic or public health policies that will affect the number of people who die in car accidents or from illness.

[5] In fact, similar impossibility results can be obtained for more general framework (i.e. simple ordinal preference domains). But I assume numerical representability of preferences for notational simplicity because it is not essential for my results.

[6] Reflexibility is defined as follows in this setting:
$\forall u_N \in U, u_N \sim u_N$.
Note that the above definition of reflexivity is equivalent to the Pareto indifferent axiom. In this sense, this setting presupposes Kaplow and Shavell's welfarism. If I assume social rankings on the set of social states or economic allocations, it is possible to derive usual welfarism (*strong neutrality*) by imposing the axioms of Pareto indifference and independence of irrelevant of alternatives (IIA). However, in a general setting of social choice or economic environment, impossibility results of this paper still survive without imposing IIA.

[7] In population ethics, this is sometimes called *betterness relations* or *population axiology*.

First, let us define two conditions that are considered to be the most basic requirement for consistency.

***Acyclicity.*** $\forall u_{N_1}, u_{N_2}, \ldots, u_{N_m} \in U, \ [u_{N_1} \succ u_{N_2}, u_{N_2} \succ u_{N_3}, \ldots, \& \ u_{N_{m-1}} \succ u_{N_m}] \rightarrow u_{N_m} \nsucc u_{N_1}$.

Acyclicity is a necessary and sufficient condition for a maximal element to exist in a finite set (Sen 1970/2017). It is a minimal requirement that must be satisfied, since a violation of acyclicity would prevent a choice of socially desirable elements (also, a cycle implies a possibility of irrational *money pump*).

***Indifference Transitivity***. $\forall u_{N_1}, u_{N_2}, u_{N_3} \in U, [u_{N_1} \sim u_{N_2} \& u_{N_2} \sim u_{N_3}] \rightarrow u_{N_1} \sim u_{N_3}$.

Indifference transitivity requires that social indifference relations are transitive. This condition may be considered too strong in some cases[8], but note that it is a much weaker condition than usual transitivity.

Next, let us define a class of the Pareto conditions. First, I introduce weak Pareto, which is a basic requirement of efficiency in the setting of the same population size.

---

[8] The example of Luce (1956) is interesting. Adding a very small amount of sugar to a cup of coffee does not change the taste significantly compared to before adding sugar. Therefore, if a person compares the taste of coffee before and after adding sugar, she will be indifferent between them. However, if this task is repeated thousands of times, indifference transitivity implies that she must be indifferent between the first cup of black coffee and the last cup of sugar-flavored coffee. This is clearly abnormal. Although transitivity is important for analytical convenience and as a sufficient condition for the existence of choice, its application imposes very strong restrictions on the identifiability of goods.

***Weak Pareto***. $\forall u_N, v_N \in U$, if $\forall i \in N, u_i > v_i$, then $u_N \succ v_N$.

This condition, also called *unanimity*, requires that if everyone strictly prefers $u_N$ to $v_N$, then society must respect this unanimous judgment.

Next efficiency axiom is a natural extension of the Pareto principle.

***Extended Pareto***. $\forall u_N, v_M \in U, \forall i \in N \cup M$, if $u_i > v_i$ or $v_i \not\succeq u_i$, then $u_N \succcurlyeq v_M$. Moreover, if $\exists j \in N \cup M, u_j > v_j$, then $u_N \succ v_M$.

This condition, also called *A-Pareto*, requires that if no one prefers $v_M$ to $u_N$ and at least one prefers $u_N$ to $v_M$, then $u_N$ is socially better than $v_M$ (Golosov et al. 2007). However, this is a very strong requirement, and it can be easily shown that this condition leads to cycles, as in the Temkin's example. Furthermore, it leads to extremely irrational judgments such as $(u + \varepsilon, \emptyset, \ldots, \emptyset) \succ (u, \ldots, u)$ for any small $\varepsilon > 0$, any positive real number $u$, and any population size $n$.[9]

Therefore, let us consider weakening this requirement to its ultimate extent. Let $n > 1$ be a fixed natural number. Suppose that in comparing profiles $u_N$ and $v_N$ with a potential population of $n$, only one individual $j$ does not prefer $v_N$ to $u_N$ (i.e. $j$ either weakly prefers $u_N$ to $v_N$ or does not exist in both of the profiles), and $n$-1 people other than $j$ strictly prefer $u_N$ to $v_N$. Then, minimal almost weak Pareto respects the almost unanimous strict preferences of $n$-1 people and requires that $u_N$ be socially preferred over $v_N$.

---

[9] If the utility level $u$ is high enough so that everyone is happy, such a decision could be considered irrational. Strictly speaking, $u$ is a just ordinary number and cannot be interpersonally comparable in this paper. Note that, however, the similar argument is easily created if $i$'s utility level $u$ is replaced by $u_i$ which means a very happy life for $i$.

***Minimal Almost Weak Pareto***. $\exists N \in \mathcal{N}, \forall u_N, v_N \in U, \text{if } \exists j \in N, v_j \not\geq u_j \ \& \ \forall i \in N \setminus \{j\}, u_i > v_i,$ then $u_N \succ v_N$.

Similarly, it is easily possible to consider an extension of Pareto indifference. Suppose that in comparing profiles $u_N$ and $v_N$ with a potential population of *n*, only one individual *j* never prefers one to another (i.e. *j* is either indifferent between them or does not exist in both of the profiles), and *n*-1 people other than *j* are indifferent between $u_N$ and $v_N$. Then, minimal almost Pareto indifference respects the almost unanimous indifferent preferences of the *n*-1 individuals and requires that $u_N$ be socially indifferent from $v_N$.

***Minimal Almost Pareto Indifference***. $\exists N \in \mathcal{N}, \forall u_N, v_N \in U, \text{if } \exists j \in N, [v_j \not\succ u_j \ \& \ u_j \not\succ v_j] \ \& \ \forall i \in N \setminus \{j\}, u_i = v_i, \text{then } u_N \sim v_N$.

There are two points worth noting about these minimal almost Pareto requirements. First, the almost Pareto judgment is only valid for a fixed number of *n* people. If the almost Pareto principle is required for any natural number *n*, transitivity implies the same problem as extended Pareto. That is, it would be possible that for any profiles in which numerous individuals enjoy the same sufficiently high utility level *u*, a profile in which only one individual enjoys slightly higher utility *u+ε* must be socially better than it.[10] Second, the almost Pareto principles say nothing about a value of a life worth living. When *i* does not exist in $u_N$ and *i*'s utility level is very low in $v_M$, it would seem that the application of the almost Pareto principles should be refrained from. However, the value of a life worth living varies considerably between individuals, so it seems to be an incomplete or ambiguous concept.[11] Note that even if the almost Pareto principles are limited to apply for such moderate cases by allowing a broad range of lives worth living, the impossibility results in this paper still survive.

---
[10] However, as I will see in the next section, this conclusion is vacuously satisfied because even the minimal almost weak Pareto principle cannot satisfy acyclicity.

## 3. Basic Results and Applications

This section provides two fundamental impossibility results about the almost Pareto principles in the previous section. First, I show that no social ranking satisfies almost weak Pareto and acyclicity.

**Theorem 1**. *There is no social ranking that satisfies minimal almost weak Pareto and acyclicity.*

Proof. Given $|N| = n$, consider the following sequence of utility profiles with the same potential individuals.

| Individual | 1 | 2 | … | $i$ | … | $n$ |
|---|---|---|---|---|---|---|
| $u_N^1 =$ | $(u_1^{2n-1},$ | $u_2^{2n-3},$ | …, | $u_i^{2n-2i+1},$ | …, | $u_n^1),$ |
| $u_N^2 =$ | $(u_1^{2n-2},$ | $u_2^{2n-4},$ | …, | $u_i^{2n-2i},$ | …, | $\emptyset),$ |
| $u_N^3 =$ | $(u_1^{2n-3},$ | $u_2^{2n-5},$ | …, | $u_i^{2n-2i-1},$ | …, | $u_n^{2n-1})$ |
| ⋮ | ⋮ | ⋮ | | ⋮ | | ⋮ |
| $u_N^{2n-2} =$ | $(u_1^2,$ | $\emptyset,$ | …, | $u_i^{2n-2i+4},$ | …, | $u_n^4),$ |

---

[11] As wrongful birth and wrongful life lawsuits, self-help groups of people with disabilities who see their own lives as precious ones, and the standards for euthanasia certification show, there are serious epistemological differences as to what standard of living is considered to be a life worth living. Hence, it is natural to think that there is a large range of thresholds due to uncertainty, incompleteness, and ambiguity in the concept of lives worth living.

| $u_N^{2n-1} =$ | $(u_1^1,$ | $u_2^{2n-1},$ | ..., | $u_i^{2n-2i+3},$ | ..., | $u_n^3),$ |
|---|---|---|---|---|---|---|
| $u_N^{2n} =$ | $(\emptyset,$ | $u_2^{2n-2},$ | ..., | $u_i^{2n-2i+2},$ | ..., | $u_n^2).$ |

Suppose that $\forall i \in N, \forall \tau \in \{2, \ldots, 2n-1\}, u_i^\tau > u_i^{\tau-1}$ in the above profiles. Then, by minimal almost weak Pareto, $u_N^1 \succ u_N^2 \succ \cdots \succ u_N^{2n}$. Also, minimal almost weak Pareto implies $u_N^{2n} \succ u_N^1$, which violates acyclicity. ∎

This result is much more serious than the Temkin example in Section 1. First, to avoid a cycle, social rankings have to grant a kind of veto power to some of *n* people, that is, the non-existence of one person must have a power to reject the almost unanimous preferences of *n*-1 people.[12] Second, since the fixed population size *n* can be any natural number, in order to avoid a cycle, a society must apply the Pareto principle only to the same population size for all population sizes. However, this restriction is too strong to compare well-being profiles in the almost contexts of social choice problems and means that the Pareto principle is desperately useless at least for intergenerational and population welfare evaluations. Third, since this result holds for social evaluations in various contexts as well, a social planner will face a fundamental impossibility result unless she completely abandons all extensions of the Pareto principle[13] or accept very strong assumptions on them as I will explain later. Fourth, the result also suggests that cycles in supermajority voting is more serious than usually expected. If ∅ is replaced with 0 in the proof of Theorem 1, it is easy to prove that a sequence of alternatives corresponding to this profile yields a cycle even in a supermajority rule that requires *n*-1 support votes.[14] In particular, when alternatives have concrete characteristics such as economic allocations, it is easy to show that such a sequence of preferences exists. Thus, the problem of cycles in supermajority voting may be much more serious than when alternatives have no mathematical structure and all preference patterns are equally probable.

---

[12] Formally, individual *j* has a veto over $\{u_N, v_N\}$ if and only if [$u_j$ and $v_j$ are noncomparable and $\forall i \in N \setminus \{j\}, u_i > v_i$] → $u_N \not\succ v_N$. Note that this definition of a veto is an extended version of the usual veto. Again, if I change the conditions so that the veto only applies to utility levels for which the comparison of non-existence and existence is incomplete, the impossibility result still holds. In fact, I can prove the impossibility theorem on incomplete domains by multiplying the sequence of increments in utility levels in the proof of Theorem 1 by a small real number *ε*.
[13] The proof of Theorem 1 shows that all the extended Pareto principles based on the partial unanimous support of the numbers of individuals (1, …, *n*-1) must violate acyclicity.

Next, I will show a problem with minimal almost Pareto indifference. This result leads to undesirable results in the sense that requiring indifference transitivity is inconsistent with weak Pareto.

**Theorem 2**. *There is no social ranking that satisfies weak Pareto, minimal almost Pareto indifference, and indifference transitivity.*

Proof. For any $u, \varepsilon > 0$, consider the following sequence of utility profiles with the same potential individuals.

| Individual | 1 | 2 | 3 | … | $n$ |
|---|---|---|---|---|---|
| $u_N^1 =$ | ($u$, | $u$, | $u$, | …, | $u$), |
| $u_N^2 =$ | ($\emptyset$, | $u$, | $u$, | …, | $u$), |
| $u_N^3 =$ | ($u + \varepsilon$, | $u$, | $u$, | …, | $u$), |
| $u_N^4 =$ | ($u + \varepsilon$, | $\emptyset$, | $u$, | …, | $u$), |
| $u_N^5 =$ | ($u + \varepsilon$, | $u + \varepsilon$, | $u$, | …, | $u$), |

---

[14] Precisely, the profile in the proof of Theorem 1 is not needed to prove a cycle. The reader could simply use the so-called *Latin square* profile and produce a cycle of the (*n*-1)-supermajority rule. The reason why I use the profile in the proof of Theorem 1 is that almost weak Pareto needs (*n*-1)-well being comparisons, which means *n* individuals must exists in either the considered profiles.

| ⋮ | ⋮ | ⋮ | | | ⋮ |
|---|---|---|---|---|---|
| $u_N^{2n} =$ | $(u + \varepsilon,$ | $u + \varepsilon,$ | $u + \varepsilon,$ | …, | $\emptyset),$ |
| $u_N^{2n+1} =$ | $(u + \varepsilon,$ | $u + \varepsilon,$ | $u + \varepsilon,$ | …, | $u + \varepsilon).$ |

By minimal almost Pareto indifference, $u_N^1 \sim u_N^2 \sim \cdots \sim u_N^{2n+1}$. By weak Pareto, $u_N^{2n+1} \succ u_N^1$ which violates indifference transitivity. ∎

This result means that even Pareto indifference should not be used in the almost comparison context. In this sense, it seems that in intergenerational evaluations, a planner should salvage this principle by significantly modifying welfare evaluations. I shall discuss the modification of the Pareto principle and the problems that come with it in the next section.

At the end of this section, let us explore that the impossibility of the almost weak Pareto principle holds in various contexts of economics. For simplicity, I consider three-dimensional comparisons in all application examples, but the following results can be easily generalized to *n*-dimensional comparisons.

**Application 1** (Aggregation with Incomplete Preferences):
Suppose there are three potential individuals with incomplete preferences. For simplicity, let us assume that the three are all highly gifted individuals who can become outstanding scholars, artisans, and artists. In terms of all the consumption bundles, honors, and human relationships that can be obtained from these occupations, none of the three individuals can compare which is better. However, let us assume that there are some differences in the number of awards and incomes in these careers, and that each career can be ranked as A⁺, A, or A⁻ (however, even A⁺-ranked occupations are not strictly better than other A⁻-ranked occupations,). In this case, the following six allocations yield a cycle of social ranking.

$u_1$ = (scholar with rank A, artisan with rank A$^+$, artist with rank A$^-$),

$u_2$ = (scholar with rank A$^-$, artisan with rank A, scholar with rank A$^+$),

$u_3$ = (artist with rank A$^+$, artisan with rank A$^-$, scholar with rank A),

$u_4$ = (artist with rank A, artist with rank A$^+$, scholar with rank A$^-$),

$u_5$ = (artist with rank A$^-$, artist with rank A, artist with rank A$^+$),

$u_6$ = (scholar with rank A$^+$, artist with rank A$^-$, artist with rank A).

Naturally, the discussion of social welfare assessment that takes into account incomplete preferences can be easily extended to the problem of welfare assessment under risky and uncertain situations when incomplete preferences are allowed. Therefore, this impossibility seems to pose a serious problem for welfare assessment in economics as a whole.[15]

**Application 2** (Aggregation of Multi-dimensional Well-being):

For simplicity, suppose that only the three dimensions of education, health, and housing are important factors in multi-dimensional well-being assessment. In terms of education, let us assume that a master's degree in electrical engineering (EE) and a master's degree in mechanical engineering (ME) cannot be compared between educational backgrounds with the top 18-20% of achievement levels. In addition, a planner cannot compare the health levels of an individual with low vision (visual acuity 0.05-0.1) and an individual with stiff neck (arm lift range 60-70 degrees). In terms of living environment, a planner cannot compare rooms (100-110 square meters) in urban areas in London and Boston. In this case, when comparing individuals with the three dimensions (education, health, and living environment) as components of well-being, a cycle occurs by almost weak Pareto comparisons as shown below.

---

[15] By considering fuzzy preferences, it is possible to artificially change them to complete preferences. There is also a methodology that takes into account a fourth relationship called *parity*. However, as will be discussed later, these methods require very large assumptions in the evaluation of individual and social welfare, and inevitably lead to the indexing dilemma problem.

$u_1 = $ (EE Top 19%, Low Vision VA 0.1, London 100 $m^2$),

$u_2 = $ (EE Top 20%, Low Vision VA 0.07, Boston 110 $m^2$),

$u_3 = $ (ME Top 18%, Low Vision VA 0.05, Boston 105 $m^2$),

$u_4 = $ (ME Top 19%, Stiff Neck Range 70°, Boston 100 $m^2$),

$u_5 = $ (ME Top 20%, Stiff Neck Range 65°, London 110 $m^2$),

$u_6 = $ (EE Top 18%, Stiff Neck Range 60°, London 105 $m^2$).

The above example is a case of intra- or inter-personal comparisons of multi-dimensional well-being, but the reader can easily make similar cycles in the context of international comparisons of multi-dimensional well-being.

**Application 3** (Intertemporal or Intergenerational Aggregation):

For this application, let's assume there are four potential individuals and two time periods, "present" and "future." Again, the following sequence of utility profiles easily yields a cycle:

|  | *Present* | | *Future* | |
| --- | --- | --- | --- | --- |
| Individual | 1 | 2 | 3 | 4 |
| $u_1 = $ | (6, | 4, | 2, | ∅), |
| $u_2 = $ | (5, | 3, | 1, | 7), |
| $u_3 = $ | (4, | 2, | ∅, | 6), |
| $u_4 = $ | (3, | 1, | 7, | 5), |
| $u_5 =$ | (2, | ∅, | 6, | 4), |
| $u_6 =$ | (1, | 7, | 5, | 3), |
| $u_7 =$ | (∅, | 6, | 4, | 2), |
| $u_8 =$ | (7, | 5, | 3, | 1). |

# 4. Solutions？

This section investigates methodologies for resolving the impossibility results about the minimal almost Pareto principles, and observes that all possible solutions either assume very strong constraints or have other logical impossibility consequences.

The first solution is to assign an objective value of a life worth living, which implicitly allows us to compare interpersonally utilities. This is the approach adopted by Blackorby and Donaldson (1984) in their formulation of critical level generalized utilitarianism.[16] However, it seems an obvious flaw that this solution implies the *weak repugnant conclusion* (Greaves 2017) and the *reverse repugnant conclusion with a threshold* (Sakamoto 2023). Suppose that there is an objective critical level for the value of individual existence. If the critical level is very low, a miserable society with a huge number of poor individuals whose utility is slightly better than the critical level ends up being socially better than a happy society with many sufficiently wealthy individuals (the weak repugnant conclusion). Conversely, if the critical level is high, a happy society with many individuals whose utility is slightly lower than the critical level but still sufficiently high ends up being socially worse than a solitude society with only one individual whose utility is at the critical level (the reverse repugnant conclusion with a threshold). Hence, the solution that uses the objective value of a life worth living must face the undesirable conclusions.[17]

---

[16] Sikora (1978) discusses a neutral level of a life worth living and proposes the Pareto-plus principle, which requires an addition of an individual above the neutral level to a utility-unaffected population to be considered as social improvement. Strictly speaking, assigning an objective value to the non-existence of an individual has two variations, fixed or variable values of non-existence. In general, the fixed value approach is based on a critical level generalized utilitarianism in Blackorby and Donaldson (1984), while the variable value approach is based on a generalization of average utilitarianism. Note that the weak sadistic conclusion must occur when a critical level of an individual's existence varies for each utility profile; see Sakamoto (2023).

The second solution is to evaluate a value of individual existence based on each individual's preference. This solution is called *P-Pareto* by Golosov et al. (2007). However, this problem also immediately runs into difficulties.[18] In fact, especially in the context of welfare evaluations of future generations, the method of assigning utility values to the non-existence of individuals raises a serious question of how to infer preferences of future generations. It is possible to set the value of individual non-existence as the expected value of possible extended utility functions that have the value of non-existence, assuming an artificially probability distribution of possible utility functions. But this is the same as assuming a de facto objective value of non-existence, so it immediately faces the same problem as the first solution (and furthermore, with this method, all potential individuals will end up having the same preference). In addition, as the Pareto improvement, this solution recommends making those who wish not to exist even though they have a high level of utility (i.e. sufficiently wealthy) into non-existence, and consisting a society with only individuals who wish to exist even if they enjoy very poor lives.[19] Hence, this concept of efficiency means that the universal set of Pareto efficient allocations would be the union of all the Pareto efficient allocations for all possible populations in which every individual enjoys more than a consumption bundle that makes her at least wish to exist. This seems to imply that this Pareto efficiency is an almost useless concept in determining the optimal population size.[20]

---

[17] Some readers might think that if the critical level were medium, there would be no problem. But even if the critical level were medium, a society with one individual whose utility is just slightly better than the critical level would be socially better than a society with a vast population in which every individual enjoys a not-so-bad life (like the so-called musak-and-potatoes life) at the critical level. In that sense, a kind of the reverse repugnant conclusion will still be alive in this context.

[18] Note that both the first and second solutions must imply the indexing dilemma, which is the problem of incompatibility between the Pareto principles and multi-dimensional Pigou-Dalton transfer in the setting of usual economic environment with a fixed population. For a comprehensive survey of the indexing dilemma, see Weymark (2017).

[19] Moreover, repeating this Pareto improvement process leads to an undesirable conclusion; an allocation in which poor individuals barely wish to exist is socially better than an allocation in which rich but arrogant individuals wish not to exist, which seems counterintuitive at least for me.

The third solution, as recommended by Sen (1970/2017), is to accept incompleteness of social rankings. However, this approach often makes it unclear how and why a pair of alternatives can be compared. In fact, the application of the Pareto principle may have to be limited to the set of alternatives with the same population, since a cycle occurs even when applying the almost weak Pareto principle. Furthermore, under the standard assumptions of weak anonymity, acyclicity, and ordinal level noncomparability, every individual must have a kind of veto, which implies alternatives that are supported by almost everyone must be indifferent or incomplete from the others.[21] That is, even if there is no interest conflict between alternatives, any anonymous and acyclic social ordering cannot be decisive over them. Such a ranking would say almost nothing in various contexts such as population assessments, multidimensional well-being aggregations, and intergenerational equity.

---

[20] In comparing $u_N$ and $v_M$, if everyone in $M$ who does not exist in $N$ wishes not to exist, everyone in $N$ who does not exist in $M$ wishes to exist, and all individuals who exist in both $N$ and $M$ prefer $u_N$ to $v_M$, then this approach says that $u_N$ is socially better than $v_M$. Even in this context, the planner would not be discussing the optimal population size. As long as an allocation is Pareto efficient in a society where all individuals want to exist, any population size will be good in the Pareto sense.

[21] Let us define weak anonymity as follows:

**Weak anonymity.** $\forall N \in \mathcal{N}, \forall$ bijections $\pi$ on $N, \forall u_N, v_N \in U, u_N \succcurlyeq v_N \leftrightarrow u_{\pi(N)} \succcurlyeq v_{\pi(N)}$.

Note that the usual anonymity axiom is too strong to compare utility profiles without imposing any interpersonal comparability of well-being. Under the usual ordinal interpersonal noncomparability, which requires that a social ranking must be invariant with respect to any monotonic transformation of utility, weak anonymity seems valid as an impartiality requirement. Then, the following result is obtained, which can be easily proven by using profiles in the proof of Theorem 1.

Theorem 3. *If a social ranking satisfies ordinal interpersonal noncomparability, weak anonymity, and acyclicity, then every individual has a veto power in the following sense:* $\forall N \in \mathcal{N}, \forall u_N, v_N \in U, \forall j \in N$, $\left[\text{if } u_j = \emptyset, v_j \in \mathbb{R}, \text{and } \forall i \in N \setminus \{j\}, u_i > v_i, \text{then } u_N \not\succ v_N\right]$ or $\left[\text{if } u_j \in \mathbb{R}, v_j = \emptyset, \text{and } \forall i \in N \setminus \{j\}, u_i > v_i, \text{then } u_N \not\succ v_N\right]$.

The last solution is to make a parent's utility function, child's utility function, and discount rate the same in the form of dynastic model, overlapping generation model, and fertility choice model, etc., which are familiar in applied economics, and to impose artificial constraints on the streams of consumption and investment bundles. Under the standard assumptions in such applied models, the domain of individual preferences is restricted to the ultimate extent, and the inconsistency problem of the Pareto principles often is ignored. This is partly because a set of optimal allocations is non-empty and efficient in these models. However, the concept of the Pareto principles is to determine which allocations is socially better than the others by respecting various individual preferences under the assumptions of ordinal and interpersonal noncomparable preferences. The claim that a set of efficient allocations is non-empty for applied models, in which individual preferences are identical, resource allocations are ultimately simplified to streams of unidimensional consumption bundles, and fertility choice often yields no conflict among future generations or is fixed exogeneously, seems to be the solution furthest from the spirit of Pareto comparisons.

## 5. Conclusion

This study considers two modified versions of the Pareto principle (minimal almost weak Pareto and minimal almost Pareto indifference), which state that when the interests of $n$-1 people in society coincide, except for one person who has no interest, society should respect consensus of almost the majority of people. Then, I show that minimal almost weak Pareto violates acyclicity and a combination of minimal almost Pareto indifference and indifference transitivity is inconsistent with weak Pareto.

Although these results are simple, they still hold in various contexts of welfare evaluations in economics. Therefore, the application of the Pareto principle needs significant limitations or modifications. If the Pareto principle is the central principle of normative economics, it is urgent to understand how strong the constraints of this principle are when considering the problem of evaluating well-being of future and current generations, and to accumulate the knowledge necessary to reboot welfare economics.

# Reference.

Blackorby, C. and D. Donaldson (1984) "Social Criteria for Evaluating Population Change," *Journal of Public Economics*, 25: 13-33.

Diamond, P. A. (1965) "The Evaluation of Infinite Utility Streams," *Econometrica*, 33(1): 170-177.

Fleurbaey, M., B. Tungodden, and H. F. Chang (2003) "Any Non-welfarist Method of Policy Assessment Violates the Pareto Principle: A Comment," *Journal of Political Economy*, 111(6): 1382-1385.

Golosov, M., L. E. Jones, and M. Tertilt (2007) "Efficiency with Endogenous Population Growth," *Econometrica*, 75(4): 1039-1071.

Greaves, H. (2017) "Population Axiology," *Philosophy Compass*, 12(11): e12442.

Kaplow, L. and S. Shavell (2001) "Any Non-welfarist Method of Policy Assessment Violates the Pareto Principle," *Journal of Political Economy*, 109 (2): 281-286.

Luce, R. D. (1956) "Semiorders and a Theory of Utility Discrimination," *Econometrica*, 24(2): 178-191.

Sakamoto, N. (2023) "How to Avoid Both the Repugnant and Sadistic Conclusions without Dropping Standard Axioms in Population Ethics," *RCNE Discussion Paper Series 11, Research Center for Normative Economics, Institute of Economic Research, Hitotsubashi University*.

Sen, A. K. (1970) "The Impossibility of a Paretian Liberal," *Journal of Political Economy*, 78(1): 152-157.

Sen, A. K. (1970/2017) *Collective Choice and Social Welfare, An Expanded Edition*, Harvard University Press, Cambridge, MA.

Sikora, R. (1978) "Is It Wrong to Prevent the Existence of Future Generations?" In *Obligations to Future Generations*, edited by R.I. Sikora and B. Barry. Philadelphia: Temple University Press.

Suzumura, K. (2001) "Pareto Principles from Inch to Ell," *Economics Letters*, 70: 95-98.

Temkin, L. S. (1987) "Intransitivity and the Mere Addition Paradox," *Philosophy and Public Affairs*, 16(2): 138-187.

Weymark, J. A. (1993) "Harsanyi's Social Aggregation Theorem and the Weak Pareto Principle," *Social Choice and Welfare*, 10: 209-221.

Weymark, J. A. (2017) "Conundrums for Nonconsequentialists," *Social Choice and Welfare*, 48(2): 269-294